\documentclass[aps,prd,nofootinbib,showpacs,superscriptaddress, preprint]{revtex4-1}
\pdfoutput=1
\usepackage[utf8]{inputenc}
\usepackage{amsmath,amssymb}
\usepackage{epsfig}
\usepackage{graphicx}
\usepackage[usenames,dvipsnames]{color}
\usepackage[colorlinks,citecolor=blue]{hyperref}
\usepackage{soul}
\usepackage{color}
\usepackage{subfigure}

\begin{document}
\title{Linking the pseudo-Dirac dark matter and radiative neutrino mass in a singlet doublet scenario}

%%%%%%%%%   Authors   %%%%%%%%%%%%
\author{Partha Konar}
\email{konar@prl.res.in}
\affiliation{Physical Research Laboratory, Ahmedabad - 380009, Gujarat, India} 
\author{Ananya Mukherjee}
\email{ananya@prl.res.in}
\affiliation{Physical Research Laboratory, Ahmedabad - 380009, Gujarat, India}
\author{Abhijit Kumar Saha}
\email{aks@prl.res.in}
\affiliation{Physical Research Laboratory, Ahmedabad - 380009, Gujarat, India}
\author{Sudipta Show}
\email{sudipta@prl.res.in}
\affiliation{Physical Research Laboratory, Ahmedabad - 380009, Gujarat, India}
\affiliation{Indian Institute of Technology, Gandhinagar - 382424, Gujarat, India}

%\preprint{\textcolor{Red}{\today}}

\begin{abstract}
We examine simple extension of the standard model with a pair of fermions, one singlet and a doublet, in a common thread linking the dark matter problem with the smallness of neutrino masses associated with several exciting features. In the presence of a small bare Majorana mass term, the singlet fermion brings in a  pseudo-Dirac dark matter capable of evading the strong spin-independent direct detection bound by suppressing the dark matter annihilation processes mediated by the neutral current. In consequence, the allowed range of mixing angle between the doublet and the singlet fermions gets enhanced substantially. Presence of the same mass term in association with singlet scalars also elevates tiny but non-zero masses radiatively for light Majorana neutrino satisfying observed oscillation data.
\end{abstract}
\maketitle

%%%%%%%%%%%%%%%%%%%%%%%%%%%%%%%%%%%%%%%%%%%%%%%%%%%%%%%%%%%%%%%%%%%%%
\section{Introduction}
%%%%%%%%%%%%%%%%%%%%%%%%%%%%%%%%%%%%%%%%%%%%%%%%%%%%%%%%%%%%%%%%%%%%%
We now boast a remarkably successful and precisely validated Standard Model (SM) of particle physics, scalar sector of which lately being examined at the Large Hadron Collider (LHC)~\cite{Aad:2014aba,Chatrchyan:2012ft}. In spite of that, many of the experimentally observed phenomena of the Universe still lacking any amicable and well-accepted explanation within this framework. One of the major mysteries of the present Universe is the fundamental nature of dark matter which has long been inferred from different celestial and cosmological observations and estimated as accounts for nearly 26$\%$ of the total energy density of the Universe. None from the trunk of SM particles owns the appropriate properties which are necessarily required to constitute a suitable candidate for cold dark matter (DM). 
Plausible origin of tiny but non-zero neutrino mass, which also unequivocally established in different solar, atmospheric and reactor neutrino oscillation experiments, remains another long-standing puzzle. Besides, questions surrounding naturalness issue, baryogenesis and dark energy persist. 
Supersymmetry~\cite{Martin:1997ns} seems to have the ability to answer many of these unresolved questions. However, lack of any clinching evidence of supersymmetry yet in LHC encourages us to build an alternative scenario beyond the Standard Model (BSM)  to explain the observed anomalies consists of dark and neutrino sectors. Although numerous proposals exist, a concrete theoretical construction of new sector that attempts to address these seemingly unrelated issues in a minimalistic manner should earn attention. 
%unified framework, which can provide solutions to multiple issues is always appealing.

In this paper, we study a simple extension of Standard Model, which offers a common origin for pseudo-Dirac dark matter interaction with the visible sector and radiative generation of neutrino mass. To look for a particle DM candidate, several dedicated direct search experiments namely XENON 1T~\cite{Aprile:2015uzo,Aprile:2017iyp}, Panda-X~\cite{Zhang:2018xdp}  {\it etc}. are ongoing. However, so far, we have not found any positive signature of DM. This hints at the possibility of DM interaction with the visible sector is weaker than the current precision of the measurements. The singlet doublet fermionic dark matter scenario is studied extensively~\cite{Yaguna:2015mva,Fiaschi:2018rky,Restrepo:2019soi,Arcadi:2018pfo,Esch:2018ccs,Calibbi:2018fqf,Maru:2017pwl,Maru:2017otg,Xiang:2017yfs,Abe:2017glm,Banerjee:2016hsk,Horiuchi:2016tqw,Calibbi:2015nha,Cheung:2013dua,Cohen:2011ec,Enberg:2007rp,DEramo:2007anh,Barman:2019aku,Barman:2019aku,DuttaBanik:2018emv,Barman:2019aku,Barman:2019tuo,Bhattacharya:2018fus,Bhattacharya:2015qpa,Bhattacharya:2017sml,Restrepo:2015ura,Freitas:2015hsa,Cynolter:2015sua,Horiuchi:2016tqw,Bhattacharya:2016lts,Bhattacharya:2016rqj,Maru:2017pwl,Wang:2018lhk,Abe:2019wku,Barman:2019oda}, and it falls within the weakly interacting massive particle (WIMP) paradigm. There are two neutral fermion states in this set up which mix with each other and the lightest one
is identified as the DM candidate. The mixing angle depends on the coupling strength of the singlet and doublet fermion with the SM Higgs. The magnitude of this mixing angle determines whether the DM is singlet like or doublet dominated. In singlet doublet model DM candidate can be probed at direct search experiments through its interaction with nucleon mediated by the SM Higgs and the neutral gauge boson. However, the null results at direct search experiments restrict the range of the mixing angle below $\lesssim 0.06$~\cite{Yaguna:2015mva}, making the DM almost purely singlet dominated.  
%Authors of Ref.~\cite{DeSimone:2010tf} (subsequently in Ref.~\cite{Narendra:2017uxl}) first proposed in a singlet fermion extended SM set up, that an impurity in the form of a small Majorana mass term for the singlet fermion in the Lagrangian would split the DM eigenstate into two non-degenerate Majorana states. 
Considering a setup where SM is extended with a singlet fermion,  Ref.~\cite{DeSimone:2010tf} (subsequently in Ref.~\cite{Narendra:2017uxl}) demonstrated that inclusion of a small Majorana mass term for the singlet fermion in the Lagrangian splits the DM eigenstate into two nearly-degenerate Majorana states with a tiny mass difference.
In the small Majorana mass limit, the splitting does not make any difference to the relic abundance analysis, however, making a vital portal to direct detection of the pseudo-Dirac DM candidate~\cite{DeSimone:2010tf}. We apply this interesting feature in the singlet doublet dark matter model by allowing a small Majorana mass term for the singlet fermion in addition to the Dirac terms for both the singlet and doublet. 
%This results in the significant elevation of the strong restriction on the singlet doublet mixing and rich DM phenomenology, which may also provide interesting implications in collider searches with rich phenomenology \cite{WorkCont}.
This inclusion brings a significant relaxation on the singlet doublet mixing angle, which is otherwise severely constrained, as discussed before.
Present model may also provide exciting implications in collider searches with rich phenomenology \cite{WorkCont}. However, it is even more appealing to note the implication in yet another sector, seemingly unrelated so far.

We make use of the same Majorana mass term for the singlet fermion in generating the low energy neutrino mass radiatively~\cite{Ma:2006km,Ma:2009gu}. The present mechanism of neutrino mass generation is also familiar as the scotogenic inverse seesaw scheme. In the process, we extend the minimal version of the singlet doublet DM framework with multiple copies of a real scalar singlet fields~\footnote{A similar exercise on the radiative generation of neutrino mass within the singlet doublet DM framework is performed in Ref.~\cite{Restrepo:2015ura} except having a pure Majorana type DM.}. These additional scalar fields can couple with the SM leptons and the doublet fermion through lepton number violating vertices. Thus in the radiative one-loop level DM particles and the singlet scalars take part in the generation of neutrino masses. As a result, the eigenvalues of the SM neutrinos are determined by the masses of DM sector particles, scalar singlets and the Majorana mass parameter of the singlet fermion. More importantly, the Majorana nature of the SM neutrino is solely determined by the introduced Majorana mass term for the singlet fermion, which also helps in successfully evading the spin-independent~(SI) constraints in dark matter. Thus the DM sector and the neutrino mass parameters are strongly correlated in the present set up which we are going to explore in detail.

The paper is organized as follows. In Section~\ref{model}, we present the structure of our model, which is primarily an extended form of the singlet doublet model. We describe the field content, their interactions and insertion of additional Majorana term. In section~\ref{darkmatter}, we discuss the consequence of our model in dark matter phenomenology. We examine the properties of our pseudo-Dirac dark matter candidate and how it extends its model parameter space evading the spin-independent direct detection limits. In Section~\ref{neutrinomass}, we explain the mechanism of radiative generation of neutrino mass and look at the parameter space where oscillation data can be satisfied simultaneously along with the dark matter constraints and relic. Finally, we conclude highlighting features of our study in Section~\ref{conclusion}.

%%%%%%%%%%%%%%%%%%%%%%%%%%%%%%%%%%%%%%%%%%%%%%%%%%%%%%%%%%%%%%%%%%%%%
\section{The Model}\label{model}
%%%%%%%%%%%%%%%%%%%%%%%%%%%%%%%%%%%%%%%%%%%%%%%%%%%%%%%%%%%%%%%%%%%%%
We extend the SM particle sector by one $SU(2)_L$ doublet fermion ($\Psi$) and one gauge singlet fermion ($\chi$). In addition, we also include three copies of a real scalar singlet field ($\phi_{1,2,3}$). The BSM fields are charged under an additional $\mathcal{Z}_2$ symmetry while SM fields transform trivially under this additionally imposed $\mathcal{Z}_2$ (see Table~\ref{tab:particle}).
%========================================
\begin{table}[t]
	\centering
	%\begin{center}
	\begin{tabular}{ |p{5cm}|p{6cm}|p{1cm}|p{1cm}|p{.6cm}| }
		\hline
		\hspace{0.8cm}BSM and SM Fields& $\hspace{0.5cm}SU(3)_C\times SU(2)_L\times U(1)_Y\equiv \mathcal{G}$ & ${U(1)}_L$& Spin &$\mathcal{Z}_2$ \\
		\hline
		\hline
		$\hspace{0.9cm}\Psi \equiv \begin{pmatrix}
		\psi^0  \\
		\psi^-
		\end{pmatrix} $  & \hspace{0.9cm}1
		\hspace{1.4cm}2 \hspace{1.3cm}-$\frac{1}{2}$\hspace{1.1cm}  &$\hspace{0.3cm}$ 0&$\hspace{0.3cm}\frac{1}{2}$& $-$
		\\
		\hline
		$\hspace{1.8cm}$ $\chi$ $ $  & {\hspace{0.9cm}1 \hspace{1.4cm}1\hspace{1.57cm}0}\hspace{1.07cm}  &$\hspace{0.3cm}$ 0&$\hspace{0.3cm}\frac{1}{2}$ & $-$\\
		\hline
		$\hspace{1cm}{\phi_i~(i=1,2,3)} $  & {\hspace{0.9cm}1 \hspace{1.4cm}1\hspace{1.57cm}0}\hspace{1cm}&$\hspace{0.3cm}$ 0 &$\hspace{0.3cm}$0 & $-$\\
		%$\hspace{1.8cm}${$N$}$ $  & {\hspace{0.9cm}1 \hspace{1.4cm}1\hspace{1.57cm}0}\hspace{1.07cm} & - \\
		\hline
		\hline
		$\hspace{0.9cm}\ell_L \equiv \begin{pmatrix}
		\nu_\ell  \\
		\ell
		\end{pmatrix} $  & \hspace{0.9cm}1
		\hspace{1.4cm}2 \hspace{1.3cm}-$\frac{1}{2}$\hspace{1.1cm}  &$\hspace{0.3cm}$ 1
		&$\hspace{0.3cm}\frac{1}{2}$ & $+$\\
		\hline
		$\hspace{0.7cm}H \equiv \begin{pmatrix}
		w^+  \\
		\frac{1}{\sqrt{2}}(v+h+iz)
		\end{pmatrix} $  & \hspace{0.9cm}1
		\hspace{1.4cm}2\hspace{1.4cm}~$\frac{1}{2}$\hspace{0.97cm}  &$\hspace{0.3cm}$ 0&$\hspace{0.3cm}$0 & $+$\\
		\hline
	\end{tabular}
	\caption{Field contents and charge assignments under the SM gauge symmetry, Lepton number, Spin and additional $\mathcal{Z}_2$.}
	\label{tab:particle}
	%\end{center}
\end{table}
The BSM fields do not carry any lepton numbers. The Lagrangian of the scalar sector is given by
\begin{align}
\mathcal{L}_{scalar}=|D^\mu H|^2+\frac{1}{2}(\partial_\mu\phi)^2-V(H,\phi),
\end{align}
where,
\begin{align}
D^\mu= \partial^\mu-ig\frac{\sigma^a}{2}W^{a\mu}-ig^\prime\frac{Y}{2}B^\mu,
\end{align}
with $g$ and $g^\prime$ being the $SU(2)_L$ and the $U(1)_Y$ gauge couplings respectively.
The scalar potential $V(H,\phi)$ takes the following form\\
\begin{align}
V(H,\phi_i)=-{\mu_H^2}\, (H^\dagger H)+\lambda_H \, (H^\dagger H)^2+\frac{\mu_{ij}^2}{2} \, \phi_i\phi_j+\lambda_{ijk}  \phi_i^2\phi_j\phi_k+\frac{\lambda_{i j}}{2} \, \phi_i \phi_j (H^\dagger H).
\end{align}
We consider $\mu_H^2,~\mu_{ij}^2$ and the quartic coupling coefficients $\lambda_{ij}$ and $\lambda_{ijk}$ are real and positive. In general the mass term for scalars ($\mu_{ij}^2$), the quartic coupling coefficients ($\lambda_{ij},~\lambda_{ijk}$)
are non diagonal. The vacuum expectation values (vev) of all the scalars $H$ and $\phi_{1,2,3}$'s after minimising the scalar potential in the limit $\mu_H^2,\mu_{ij}^2>0$ are obtained as,
\begin{align}\label{eq:vac}
\langle H\rangle=v,~~\langle\phi_{1,2,3}\rangle=0.
\end{align} 
 Since all the quartic couplings are positive, the scalar potential is bounded from below in any field direction with the set of stable vacuum in Eq.(\ref{eq:vac}) \cite{Kannike:2012pe,Chakrabortty:2013mha}. For sake of simplicity \footnote{In the present analysis the quartic couplings for the singlet scalars have negligible role and can take any arbitrary positive value within their respective perturbativity bounds \cite{Horejsi:2005da,Bhattacharyya:2015nca}.} we assume that $\mu_{ij}^2$, $\lambda_{ij}$,~$\lambda_{ijk}$ are diagonal with the masses of the scalar fields parametrised as ($M_{\phi_1}^2,M_{\phi_2}^2,M_{\phi_3}^2$). The discrete symmetry $\mathcal{Z}_2$ remains unbroken  since $\langle\phi_{1,2,3}\rangle=0$.  
The Lagrangian for the fermionic sector (consistent with the charge assignments) is written as:
\begin{align}
\mathcal{L}=\mathcal{L}_{f}+\mathcal{L}_{Y},
\end{align}
where,
\begin{align}
\mathcal{L}_{f}=& \;  i\overline{\Psi}_L \gamma_\mu D^\mu \Psi_L+i\overline{\Psi}_R \gamma_\mu D^\mu \Psi_R+i\overline{\chi}_L  \gamma_\mu\partial^\mu \chi_L+i\overline{\chi}_R  \gamma_\mu\partial^\mu \chi_R  \nonumber\\
&-M_\Psi\overline{\Psi}_L \Psi_R-M_\Psi\overline{\Psi}_R \Psi_L  %\nonumber\\
-M_\chi \overline{\chi}_L \chi_R-\frac{m_{\chi_L}}{2}\overline{\chi^c_L}\chi_L-\frac{m_{\chi_R}}{2}\overline{\chi^c_R} \chi_R,\label{eqn:FeynA}
\end{align}
and
\begin{align}
\label{eqn:Fyukawa}
\mathcal{L}_{Y}=Y\overline{\Psi}_L \tilde{H} \chi_R+h_{ij} \overline{\ell_i} \Psi_R \phi_j+h.c..
\end{align}
We keep a small Majorana mass ($m_{\chi_{L,R}}\ll M_\chi$) term for the $\chi$ field in Eq.~(\ref{eqn:FeynA}). In this particular set up the lightest neutral fermion is a viable dark matter candidate which has a pseudo-Dirac nature provided a tiny $m_{\chi_{L,R}}$ exists. The choice of this non-vanishing $m_{\chi_{L,R}}$ is kept from the necessity of evading strong spin-independent dark matter direct detection bound. As we will see later that this term is also helpful in generating light neutrino mass radiatively. The first term in Eq.~(\ref{eqn:Fyukawa}) provides the interaction of DM with the SM particles mediated through the Higgs. While the second term in Eq.~(\ref{eqn:Fyukawa}) violates the lepton number explicitly~\footnote{Consideration of complex scalar singlets instead of real ones would lead to the conservation of the lepton number~\cite{Restrepo:2015ura}.}. This kind of lepton number violation could trigger a thermal or non-thermal leptogenesis (baryogenesis) in the early Universe, provided sufficient CP asymmetry is generated~\cite{WorkCont}.

%%%%%%%%%%%%%%%%%%%%%%%%%%%%%%%%%%%%%%%%%%%%%%%%%%%%%%%%%%%%%%%%%%%%%
\section{Dark Matter}\label{darkmatter}
%%%%%%%%%%%%%%%%%%%%%%%%%%%%%%%%%%%%%%%%%%%%%%%%%%%%%%%%%%%%%%%%%%%%% 
The different variants of singlet doublet fermion dark matter are extensively studied in the literature~\cite{Yaguna:2015mva,DuttaBanik:2018emv,Barman:2019aku,Barman:2019tuo,Bhattacharya:2018fus,Bhattacharya:2015qpa,Bhattacharya:2017sml,Restrepo:2015ura,Fiaschi:2018rky,Restrepo:2019soi,Arcadi:2018pfo,Esch:2018ccs,Calibbi:2018fqf,Maru:2017pwl,Maru:2017otg,Xiang:2017yfs,Abe:2017glm,Banerjee:2016hsk,Horiuchi:2016tqw,Calibbi:2015nha,Cheung:2013dua,Cohen:2011ec,Enberg:2007rp,DEramo:2007anh} over the years. Here we go through the DM phenomenology in brief. In the present study, we consider $M_\phi \gg M_\psi, m_{\chi_{L,R}}$ such that the role $\phi$ fields in DM phenomenology is minimal \footnote{In principle, scalars could take part in DM phenomenology through coannihilation processes. However, considering the mass pattern, we have chosen for simplicity, their contributions turn out to be negligible.}. The Dirac mass matrix for the neutral DM sector after the spontaneous breakdown of the electroweak symmetry is obtained as (in $m_{\chi_{L,R}}\rightarrow 0$ limit),
\begin{align}
 \mathcal{M}_D=\begin{pmatrix}
		M_\Psi & M_D  \\
		M_D & M_\chi 
		\end{pmatrix},
\end{align}
where we define $M_D=\frac{Yv}{\sqrt{2}}$. Therefore, we are left with two neutral Dirac particles which we identify as $(\xi_1,\xi_2)$. The mass eigenvalues of $(\xi_1,\xi_2)$ are given by,
\begin{align}\label{massE}
&M_{\xi_1} \approx M_\chi - \frac{M_D^2}{M_{\Psi}-M_\chi}\\
&M_{\xi_2} \approx M_\Psi + \frac{M_D^2}{M_{\Psi}-M_\chi}
\end{align} 
%========================================
\begin{figure}[t]
\centering
\includegraphics[scale=.5]{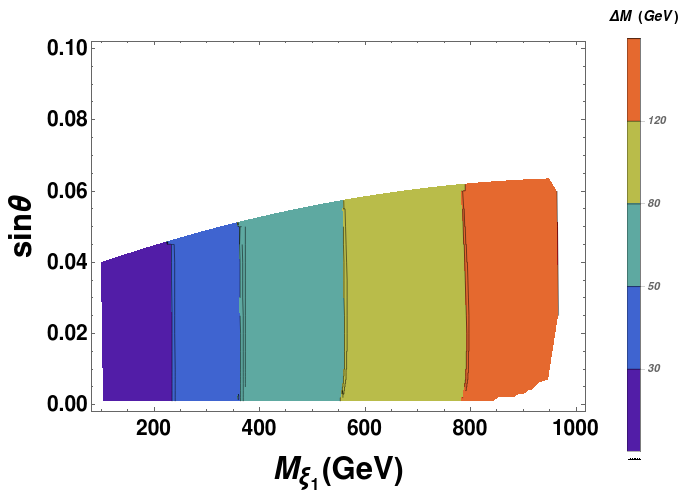}\\
\caption{Region of parameter space allowed from both the relic density and direct detection bounds are shown in a plane of  dark matter mass $M_{\xi_1}$ and mixing angle $\sin\theta$, in the limit Majorana mass $m_{\chi_{L,R}} = 0$.  Different colors are for different values of mass gap $\Delta M = (M_{\xi_2} - M_{\xi_1})$ allowed here. In this scenario, upper limit in $\sin\theta$ is strongly  constrained from direct detection bounds which gradually relaxed with higher dark matter mass and thus a lower cross section.}
\label{fig:DM1} 
\end{figure}
%========================================
Therefore, the lightest state is $\xi_1$, which we identify as our DM candidate. The DM stability is achieved by the unbroken $\mathcal{Z}_2$ symmetry. The mixing between two flavor states, {\it i.e.} neutral part of the doublet ($\psi^0$) and the singlet field ($\chi$) is parameterised by $\theta$ as
\begin{align}
\sin2\theta\simeq \frac{2Yv}{\Delta M},
\end{align}
where $\Delta M =M_{\xi_2}-M_{\xi_1}\approx M_\Psi-M_\chi$ in the small $Y$ limit.
In small mixing case, $\xi_1$ can be identified with the singlet $\chi$. The DM phenomenology is mainly controlled by the following independent parameters.
\begin{align}
\{M_{\Psi},~M_{\chi},~\theta\}.
\end{align}  

The DM would have both annihilation and coannihilation channels to SM particles, including the gauge bosons~\cite{DEramo:2007anh,Calibbi:2015nha}. It turns out that the coannihilation channels play the dominant role in determining the relic abundance for pure singlet doublet fermion DM since the annihilation processes are proportional to the square of mixing angle and hence suppressed in the small mixing limit. The DM can be searched directly through its spin-independent scattering with nucleon mediated by both SM Higgs and Z boson. 
In Fig.~\ref{fig:DM1} we show the observed relic abundance by Planck 2018~\cite{Aghanim:2018eyx} and spin-independent direct detection bounds (from XENON 1T~\cite{Aprile:2017iyp}) satisfied region in $\sin\theta-M_{\xi_1}$ plane for different values of $M_{\xi_2}$ in the absence of the Majorana mass term ($m_{\chi_{L,R}}$). We have  used \textsf{Micromega 4.3.5}~\cite{Barducci:2016pcb} package for the numerical analysis. It is observed that the relic abundance is satisfied for a particular $M_{\xi_1}$ when $\Delta M=M_{\xi_2}-M_{\xi_1}$ is small. This means the coannihilation processes are dominant compared to the annihilation processes in determining the observed relic abundance. One important point to note is that the required amount of $\Delta M$ increases with the DM mass for any fixed value of $\sin\theta$. Fig.~\ref{fig:DM1} also evinces strong constraint on $\sin\theta \lesssim 0.06$ primarily from the direct detection bounds, which gradually relaxed with higher dark matter masses because of a lower cross section. Finally, it keeps the DM framework alive from spin-independent direct detection bound. 

%========================================
\begin{figure}[t]
\centering
\includegraphics[scale=1.3]{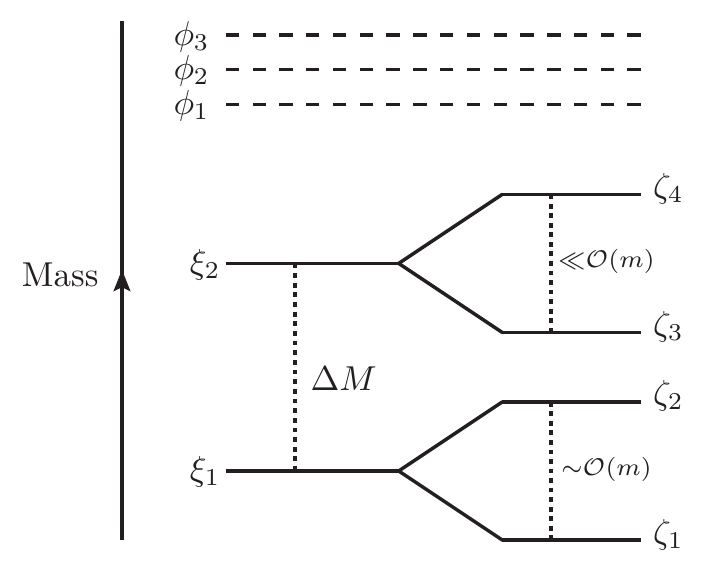}
\caption{Mass spectrum of the dark sector, showing the lightest pseudo-Dirac mode as dark matter and other heavy BSM fermions and scalars. Generation of large mass difference ($\Delta M$) and small mass gap ($m$) discussed at the text expressed at the zeroth order of $\delta_r$. Scalars are assumed to be heavier in this study.}
\label{massorder}
\end{figure}
%========================================

The strong upper bound on $\sin\theta$ can be alleviated by taking the presence of $m_{\chi_{L,R}}$ into account. The tiny nature of $m_{\chi_{L,R}}$
makes $\xi_1$ pseudo-Dirac. In the limit $m\rightarrow 0$ where we define $m=(m_{\chi_L}+m_{\chi_R})/2$, the Majorana eigenstates of $\xi_1$ ({\it i.e.} $\zeta_1,~\zeta_2$) become degenerate. The presence of a non-zero $m_{\chi_{L,R}}$ breaks this degeneracy, and we can still write
\begin{align}
&\zeta_1\simeq \frac{i}{\sqrt{2}}(\xi_1-\xi_1^c),\\
&\zeta_2\simeq \frac{1}{\sqrt{2}}(\xi_1+\xi_1^c).
\end{align}
in the pseudo-Dirac limit $m\ll M_{\zeta_1},M_{\zeta_2}$ where $M_{\zeta_1,\zeta_2}\simeq M_{\xi_1}\mp m$. Similarly, the state $\xi_2$ is spilt into $\zeta_3$ and $\zeta_4$. Hence we will have four neutral pseudo-Dirac mass eigenstates in the DM sector. The complete mass spectrum of the neutral dark sector particles is displayed in Fig.~\ref{massorder}. The mass of the charged fermion $\psi^-$
lies in between $\zeta_3$ and $\zeta_2$ as followed from Eq.~(\ref{massE}).
The pseudo-Dirac nature of the eigenstates forbid the interaction of DM ($\zeta_1$) with the neutral current mediated by SM $Z$ boson at zeroth order of $\delta_r\simeq(m_{\chi_L}-m_{\chi_R})/m_{\xi_1}$.
Thus the pseudo-Dirac DM could have the potential to escape the SI direct search bound. Although at next to leading order, the DM still possesses  non-vanishing interaction with $Z$ boson depending on the magnitude of $\delta_r$. This is analyzed in the next paragraph. It is important to note that the $m$ can not be arbitrarily small since there exists a possibility of the lighter state $\zeta_1$ to scatter inelastically with the nucleon to produce heavier state $\zeta_2$~\cite{Cui:2009xq,Hall:1997ah,TuckerSmith:2001hy}. It imposes some sort of lower bound on $m \gtrsim \mathcal{O}(1)$ KeV~\cite{Cui:2009xq,Hall:1997ah,TuckerSmith:2001hy} in order to switch off such kind of interaction. However,  the presence of a vertex like $\bar{\zeta_1}\gamma^\mu\zeta_2$ can give rise to huge $Z$ mediated s-channel coannihilation cross section of the DM with the next to lightest state (NLSP)~\cite{Hall:1997ah}  in the above mentioned limiting value of $m$. This cross section would have a suppression factor of $\sin^4\theta$. In spite of this, for moderate values of $\sin\theta$, the cross section can turn huge. We have examined and found that keeping $m\sim \mathcal{O}(1)$ GeV effectively prevents the $Z$ mediated s-channel coannihilation of the DM with the NLSP~\cite{TuckerSmith:2001hy} even with moderate values of $\sin\theta$.
A similar result is obtained in Ref.~\cite{DeSimone:2010tf,Davoli:2017swj}. At linear order in $\delta_r$, a direct search of pseudo-Dirac dark matter through Z-mediation is still possible which we discuss below.

%========================================
\begin{figure}[t]
\centering
\includegraphics[scale=.5]{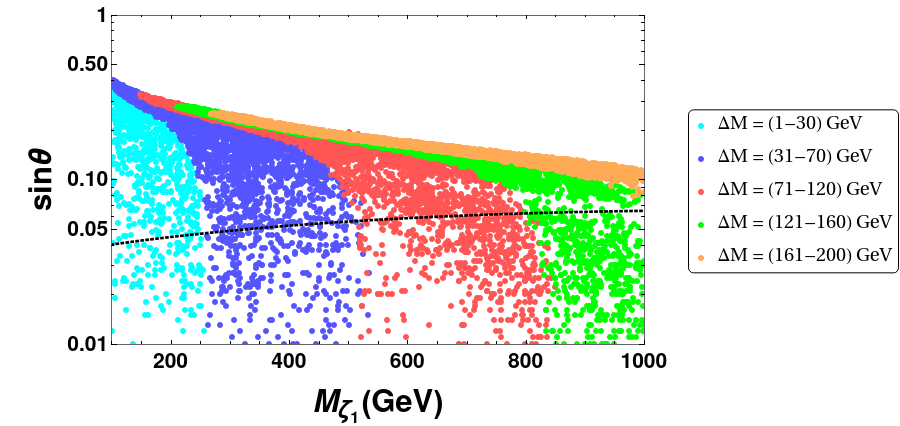}\\
\caption{Region of parameter space allowed from both the relic density and direct detection bounds are shown in a plane of dark matter mass $M_{\zeta_1}$ and mixing angle $\sin\theta$, in case of a nonzero but small Majorana mass $m_{\chi_{L,R}}$ insertion.  Different colors are for different values of mass gap $\Delta M = (M_{\xi_2} - M_{\xi_1})$ allowed here. It is instructive to compare this present plot with Fig.\ref{fig:DM1}. Unlike the previous $m_{\chi_{L,R}} =0$ case (denoted by black dotted line here), upper limit from direct detection is much relaxed and barely constrained in this scenario. The present upper limit in $\sin\theta$ is primarily constrained from the relic density criteria and (unlike the previous case) constrain is being stronger at higher dark matter mass.
}
\label{Fig2}
\end{figure}
%========================================

The vector operator for the SI direct search process mediated by $Z$ boson
will be modified to 
\begin{align}
\mathcal{L}\supset\alpha(\bar{\zeta_1}\gamma^\mu\zeta_1)(\bar{q}\gamma_\mu q),
\end{align}
with $\alpha=\frac{4 g^2\delta_r\sin^2\theta }{ m_Z^2\cos^2\theta_W}C_V^q=\alpha^\prime C_V^q$ and $g$ as the $SU(2)_L$ gauge coupling constant. Note that, at zeroth order in $\delta_r$, vector boson interaction of dark matter would vanish, and only the Higgs mediated processes would contribute to the direct search. 
Considering DM mass larger than the nucleon mass, the spin-independent direct detection cross section per nucleon is obtained as~\cite{Yaguna:2015mva,Restrepo:2019soi}
\begin{align}
\sigma^{\rm SI}\simeq \frac{a}{\pi}\frac{M_{\zeta_1}^2m_N^2\alpha^{\prime^2}}{(M_{\xi_1}+m_N)^2A^2}\Big[ZC_V^p+(A-Z)C_V^n\Big]^2,
\end{align}
where $m_N=940$ MeV, the nucleon mass, $\theta_W$ is the Weinberg angle and $C_V^p=\frac{1}{2}(1-4\sin^2\theta_W)$,~$C_V^n=-\frac{1}{2}$. It is clear from the smallness of the term $(1-4\sin^2\theta_W)$ that, the DM particle rarely talks to protons, and hence the SI cross section mainly depends on the DM interaction with neutrons. For Dirac fermion $a=1$~\cite{Halzen:1984mc}, while for Majorana $a=\frac{1}{4}$~\cite{Halzen:1984mc}. From the above relation, one can extract $\delta_r$ as follows,
\begin{align}
\delta_r=1.07\times 10^{19}\left(\frac{\sigma^{\rm SI}}{\rm cm^2}\right)^{1/2}\left(\frac{1}{\sin^2\theta}\right).
\label{eq:delta1}
\end{align}
Now to evade direct search constraints for the DM mass $\gtrsim 100$ GeV, it is sufficient to have $\sigma^{\rm SI}\lesssim 10^{-47}~{\rm cm}^2$. Imposing this bound in Eq.~(\ref{eq:delta1}), we can report an upper bound on the difference of Majorana mass parameters $m_{\chi_L}-m_{\chi_R}$ which is,
\begin{align}
m_{\chi_L}-m_{\chi_R} \lesssim 3.4\times 10^{-5}\frac{M_{\zeta_1}}{\sin^2\theta}.
\end{align}
The above bound turns out to be strongest for smaller $M_{\zeta_1}$ and larger $\sin\theta$. For the present analysis, where we accommodate a WIMP like candidate with mass $\mathcal{O}(100)$ GeV and $\sin\theta\lesssim 0.3$. This automatically sets the bound as follows
\begin{align}\label{mchibound}
m_{\chi_L}-m_{\chi_R} \lesssim 13.5 {\rm ~MeV}.
\end{align}

Taking the contribution of the Z mediated interaction of the DM with nucleon of the order of $\mathcal{O}(10^{-47}){\rm~ cm}^2$ and considering $m_{\chi_L}\simeq m_{\chi_R}=1$ GeV, we have plotted the relic abundance and direct search allowed points on $ \sin \theta - M_{\zeta_1}$ plane in Fig.~\ref{Fig2}. Different colors are presented for different values of mass gap $\Delta M = (M_{\xi_2} - M_{\xi_1})$ allowed here. It is instructive to compare this present plot with Fig.~\ref{fig:DM1}. Unlike the previous $m_{\chi_{L,R}} =0$ case (upper constraint limit of which is illustrated by a black dotted line in current plot), here upper limit from direct detection is much relaxed and barely constrains this scenario. In fact, the present upper limit in $\sin\theta$ is primarily constrained from the relic density criteria, and unlike the previous case, the constraint is being stronger at higher dark matter mass. From this analysis, it is clear that the earlier obtained limit on $\sin \theta$ got relaxed at a considerably good amount. Another notable feature of Fig.~\ref{Fig2} is that for lighter DM, large mass splitting is allowed for higher values of $\sin\theta$. This follows from the fact that the annihilation cross section starts to play an equivalent role as coannihilation at large $\sin\theta$. The above values of Majorana mass parameters would be used to evaluate the neutrino mass.

The allowed parameter space of DM in Fig.~\ref{Fig2} is also subject to indirect detection constraints. The indirect search for dark matter experiments aims to detect the SM particles produced through DM annihilation in a different region of our observable universe where DM is possibly present abundantly, such as  the  center of our galaxy or satellite galaxies. Among the many final states, photon and neutrinos, being neutral and stable can reach the indirect detection experiments without significant deviation in the intermediate regions. Strong constraint is deduced from the measured photons at space based telescopes like the Fermi-LAT or ground based telescopes like MAGIC~\cite{Ahnen:2016qkx}. The photon flux in a specific energy range is written as 
%\begin{align}\label{eq:Photon}
% \Phi_F=\frac{1}{4\pi}\frac{\langle\sigma v\rangle_{\rm ann}}{2m_{DM}^2}\int_{E_{\rm min}}^{E_{\rm max}}\frac{dN_\gamma}{dE_\gamma}dE_\gamma \int_{\rm LOS} dx \rho^2(r(b,l,x)),
%\end{align}
%where $r(b,l,x)$ is the distance of the DM halos from the galactic center. Galactic coordinates are represented by $b,l$ and $\rho(r)$ is the DM density profile. The Eq.(\ref{eq:Photon}) is rewritten as
\begin{align}
 \Phi_F=\frac{1}{4\pi}\frac{\langle\sigma v\rangle_{\rm ann}}{2m_{DM}^2}\int_{E_{\rm min}}^{E_{\rm max}}\frac{dN_\gamma}{dE_\gamma}dE_\gamma\times J,
 \label{eq:PFlux}
\end{align}
where $J=\int dx \rho^2(r(b,l,x))$ encapsulate the cosmological factors, conventionally known as $J-$factor, representing the integrated DM density within the observable solid angle along the line of sight (${\rm LOS}$) of the location. $r(b,l,x)$ is the distance of the DM halo in coordinate represented by $b,l$ and $\rho(r)$ is the DM density profile. 
%and ${\rm LOS}$ stands for line of sight.
From the observed Gamma ray flux produced by DM annihilations,  one can restrict the relevant parameters which contribute to the  DM annihilation into different charged final states like $\mu^+\mu^-$, $\tau^+\tau^-$, $W^+W^-$ and $b^+b^-$.

%========================================
\begin{figure}[t]
\centering
\includegraphics[height=8cm,width=14cm]{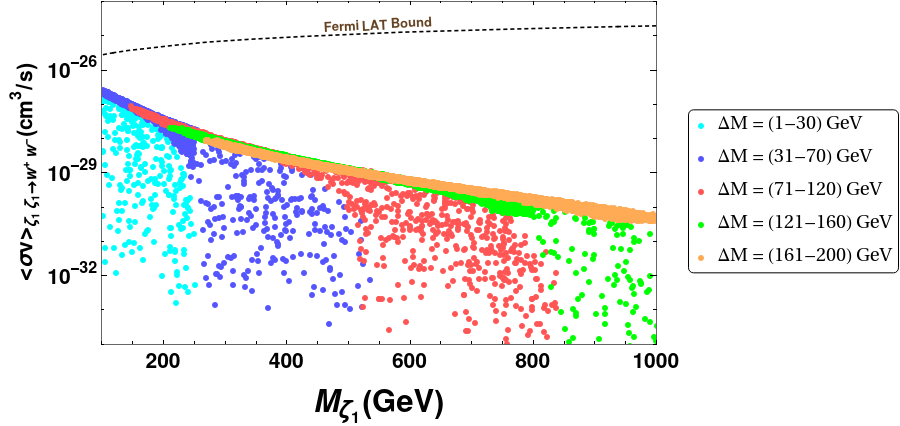}
\caption{Annihilation cross sections for relic and direct search satisfied points of DM (see Fig. \ref{Fig2}) to $W^+W^-$ final states for different sets of $\Delta M$. The bound from Fermi LAT+MAGIC~\cite{Ahnen:2016qkx} is also included for comparison purpose.}
\label{fig:InW}
\end{figure} 
%========================================

Let us recall that the relic satisfied region in Fig. \ref{Fig2} is mostly due to the coannihilation effects provided the DM annihilations remain subdominant. Although for larger $\sin\theta$, DM annihilations start to contribute to the relic density at a decent amount.
Among the many final states of DM annihilation in our scenario, $\langle \sigma v\rangle_{\zeta_1\zeta_1}$ is the dominant one with contributions from both s and t channels mediated by $\psi^\pm$ and the SM Higgs. In particular, the annihilation channels having $W^\pm$ in the final states involve $SU(2)_L$ gauge coupling. Therefore, to check the consistency of our framework against the indirect detection bounds, we focus on DM annihilation into W-pair ${\zeta_1\zeta_1\rightarrow W^+W^-}$. 
In Fig. \ref{fig:InW}, we exhibit the magnitude of $\langle\sigma v\rangle_{\zeta_1\zeta_1\rightarrow W^+W^-}$ for all the relic satisfied points in Fig. \ref{Fig2} and compare it with the existing experimental bound from Fermi-Lat~\cite{Ahnen:2016qkx}. We see that all the relic satisfied points lie well below the experimental limit. We also confirm that the model precisely satisfies the indirect search bounds on other relevant final state charged particles.

Before we end this section, it is pertinent to note that in this analysis, our focus was on the DM having mass in between hundred GeV to one TeV. Naturally, a question emerges that what happens for the higher DM masses. Since we have two independent parameters, namely $\Delta M$ and $\sin\theta$, it is possible to account for the correct order of relic abundance for any arbitrary DM mass by tuning one of these. Besides, stringent direct search bound can also be escaped easily with a vanishing tree level neutral current (due to pseudo-Dirac nature of DM) unless $\sin\theta$ turns extremely large. We have numerically checked that even for DM as massive as 50 TeV, both relic density and direct search constraints can be satisfied in the present framework.
However, a model independent conservative upper-bound on WIMP DM mass can be drawn using partial-wave unitarity criteria. The analysis performed in~\cite{Griest:1989wd} points out that a stable elementary particle produced from thermal bath in the early Universe can not be arbitrarily massive (~$\lesssim 34$ TeV~) corresponding to $\Omega h^2\sim 0.1$. Since it is a model independent bound, it applies in our case too. 
%%%%%%%%%%%%%%%%%%%%%%%%%%%%%%%%%%%%%%%%%%%%%%%%%%%%%%%%%%%%%%%%%%%%%
\section{Neutrino Mass}\label{neutrinomass}
%%%%%%%%%%%%%%%%%%%%%%%%%%%%%%%%%%%%%%%%%%%%%%%%%%%%%%%%%%%%%%%%%%%%%

\begin{figure}
\centering
\includegraphics[scale=1]{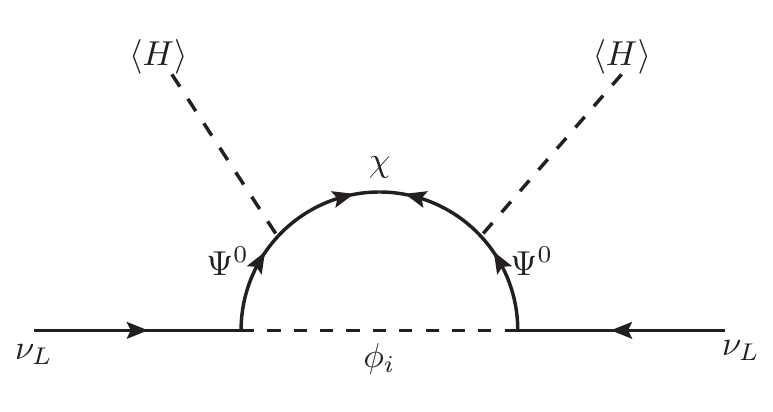}\\
\caption{Generation of neutrino mass radiatively at one loop level getting contributions from tiny Majorana mass term inserted in the dark sector along with the heavy singlet scalars.}
\label{fig:Neu_mass}
\end{figure}

In the presence of the small Majorana mass term ($m_{\chi_{L,R}}$) of $\chi$  field and the lepton number violating operator in Eq.~(\ref{eqn:Fyukawa}), it is possible to generate active neutrino mass radiatively at one loop as displayed in Fig.~\ref{fig:Neu_mass}. It is worth mentioning that this type of mass generation scheme is known as one loop generation of inverse seesaw neutrino mass~\cite{Fraser:2014yha}. 

%========================================
%========================================

The neutrino mass takes the form as provided below~\cite{Ma:2006km,Ma:2009gu,Fraser:2014yha},
\begin{align}\label{mnu}
m_{\nu_{ij}}=h^T_{ki} \Lambda_{kk} h_{jk}, 
\end{align}
where, 
$\Lambda_{kk}=\Lambda_{kk}^L+\Lambda_{kk}^R$ with
\begin{align}
\Lambda_{kk}^L=m_{\chi_L} \cos^2\theta\sin^2\theta\Big[&
\int\frac{d^4q}{(2\pi)^4}\frac{M_{\xi_1}^2}{(q^2-M_{\phi_k}^2)(q^2-M_{\xi_1}^2)^2}
+\int\frac{d^4q}{(2\pi)^4}\frac{M_{\xi_2}^2}{(q^2-M_{\phi_k}^2)(q^2-M_{\xi_2}^2)^2}\nonumber\\
&-\int\frac{d^4q}{(2\pi)^4}\frac{2M_{\xi_1}M_{\xi_2}}{(q^2-M_{\phi_k}^2)(q^2-M_{\xi_1}^2)(q^2-M_{\xi_2}^2)}\Big],
\end{align}
and
\begin{align}
 \Lambda_{kk}^R=m_{\chi_R} \cos^2\theta\sin^2\theta\Big[&
 \int \frac{d^4q}{(2\pi)^4}\frac{q^2}{(q^2-M_{\phi_k}^2)(q^2-M_{\xi_1}^2)^2}+\int \frac{d^4q}{(2\pi)^4}\frac{q^2}{(q^2-M_{\phi_k}^2)(q^2-M_{\xi_2}^2)^2}\nonumber\\
 &-\int \frac{d^4q}{(2\pi)^4}\frac{2q^2}{(q^2-M_{\phi_k}^2)(q^2-M_{\xi_1}^2)(q^2-M_{\xi_2}^2)}\Big]
\end{align}
The $h_{ij}$ is the Yukawa coupling as defined in Eq.~(\ref{eqn:Fyukawa}). 
Each integral of the above two expressions for $\Lambda_{kk}$ can be decomposed as two 2-point \textsf{Passarino-Veltman} 
 functions~\cite{tHooft:1978jhc,Ellis:2007qk} as provided below: 
\begin{align}
 \Lambda_{kk}^L= \frac{1}{16\pi^2}m_{\chi_L}\cos^2\theta \sin^2\theta\Big[&
 \frac{M_{\xi_1}^2}{M_{\phi_k}^2 - M_{\xi_1}^2} \{B(0, M_{\xi_1},M_{\phi_k}) - B(0, M_{\xi_1},M_{\xi_1})\} \nonumber\\
 &+ \frac{M_{\xi_2}^2}{M_{\phi_k}^2 - M_{\xi_2}^2} \{B(0, M_{\xi_2},M_{\phi_k}) - B(0, M_{\xi_2},M_{\xi_2})\}\nonumber\\ 
 &-  \frac{2 M_{\xi_1}M_{\xi_2}}{M_{\xi_2}^2-M_{\xi_1}^2} \{B(0, M_{\xi_2},M_{\phi_k}) - B(0, M_{\xi_1},M_{\phi_k})\}\Big], \\
%\end{align}
%\begin{align}
 \Lambda_{kk}^R= \frac{1}{16\pi^2}m_{\chi_R}\cos^2\theta \sin^2\theta\Bigg[&
 \{B(0, M_{\xi_1},M_{\phi_k})-B(0, M_{\xi_2},M_{\phi_k})\}\nonumber\\  &
 \left\{1+\frac{2M_{\xi_1}}{M_{\xi_2}^2-M_{\xi_1}^2}(M_{\xi_1}-\frac{m_{\chi_L}}{m_{\chi_R}}M_{\xi_2})\right\}\Bigg]+\frac{m_{\chi_L}}{m_{\chi_R}}\Lambda_{kk}^L
,
\end{align}
where $B(p, m_1, m_2)$ is defined as~\cite{Abe:2018emu},
\begin{align}
B(p,m_1,m_2)&= \int_{0}^1 dx \Big[ \frac{2}{\tilde \epsilon} + \text{log}\Big(\frac{\mu^2}{{m_1^2 \, x} + m_2^2 \, (1-x) - p^2 \, x \, (1-x)}\Big)\Big],
\end{align}
with, $\frac{2}{\tilde \epsilon} = \frac{2}{\epsilon} - \gamma_E + \text{log} (4 \pi)$,~$\epsilon = n-4 $ and $\gamma_E$ is the Euler-Mascheroni constant.

The mass scale $\Lambda_{kk}$ is a function of DM mass, mixing angle $\theta$ and the masses of the scalar fields. The pseudo Dirac DM phenomenology restricts $\sin\theta$ for a particluar DM mass in order to satisfy both relic and direct detection bound. Using that information one can estimate $\Lambda_{kk}$ for both higher and lower values of $\sin\theta$ for a particular DM mass. We use \textsf{QCDloop}~\cite{Ellis:2007qk} to evaluate $\Lambda_{kk}$ numerically and which is found to be consistent with the analytical estimation of $\Lambda_{kk}$.  
%========================================
\begin{figure}[t]
\centering
\includegraphics[scale=0.35]{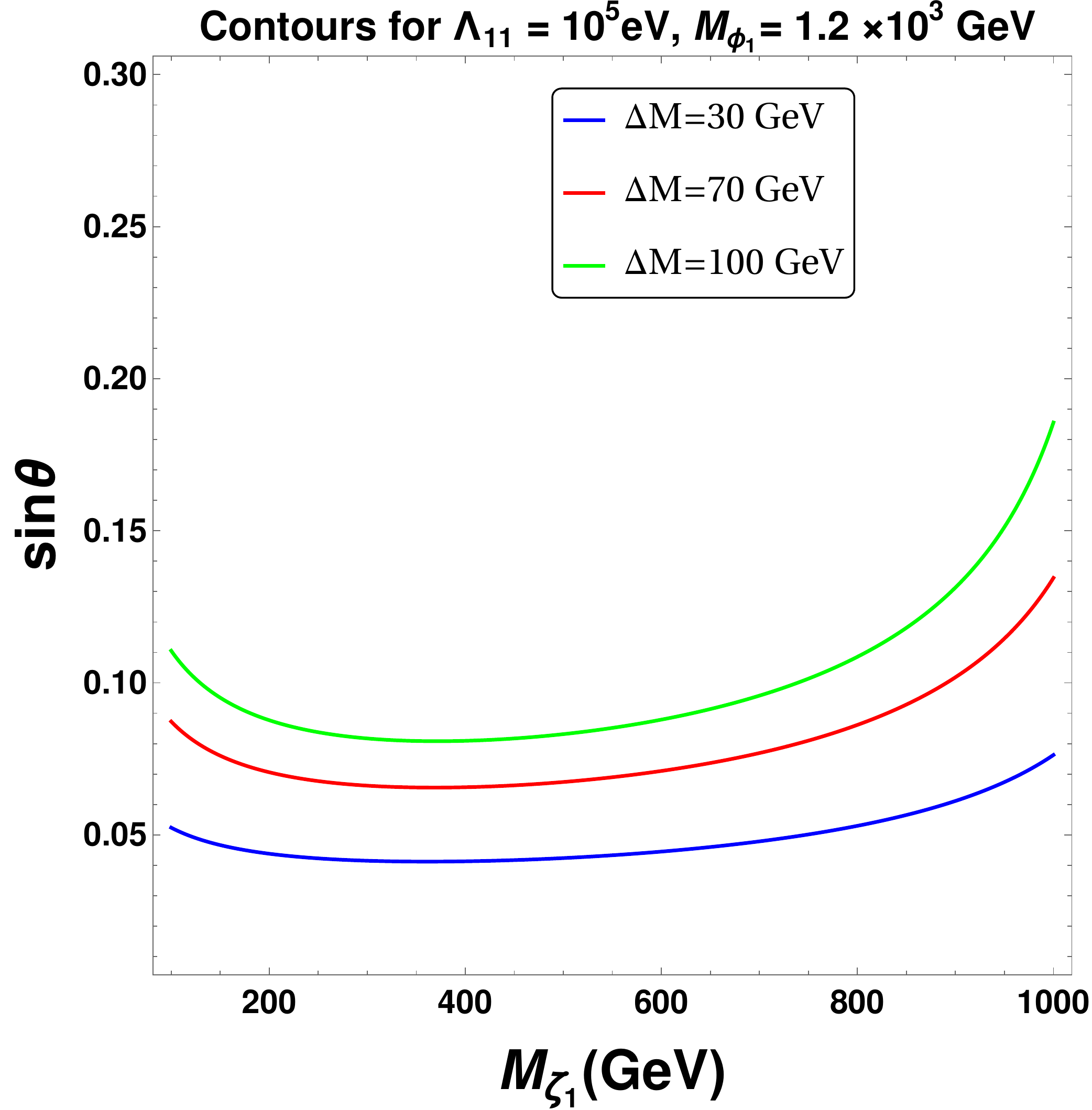}~
\includegraphics[scale=0.35]{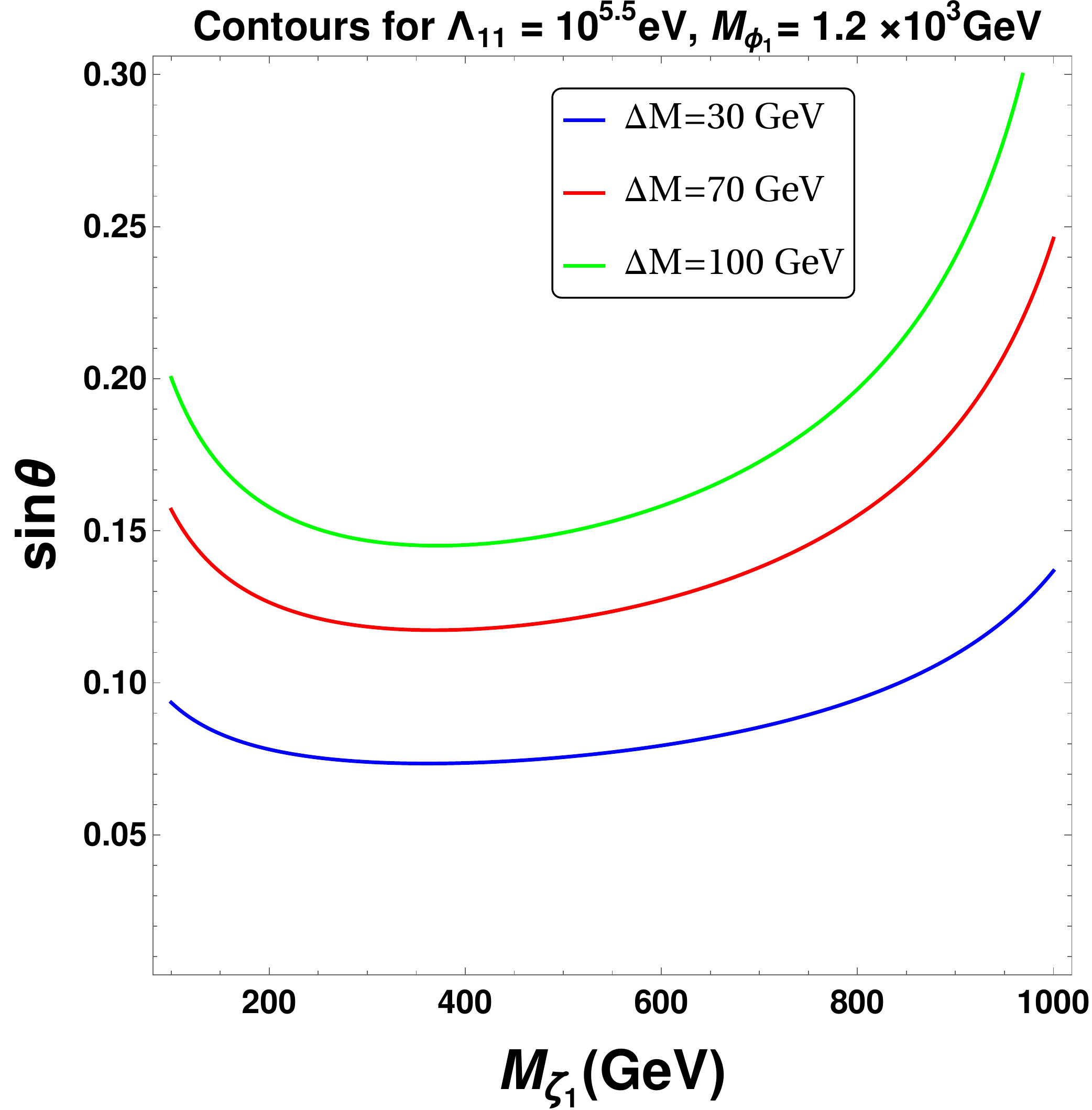}\\  \vspace{0.2 cm}
\includegraphics[scale=0.35]{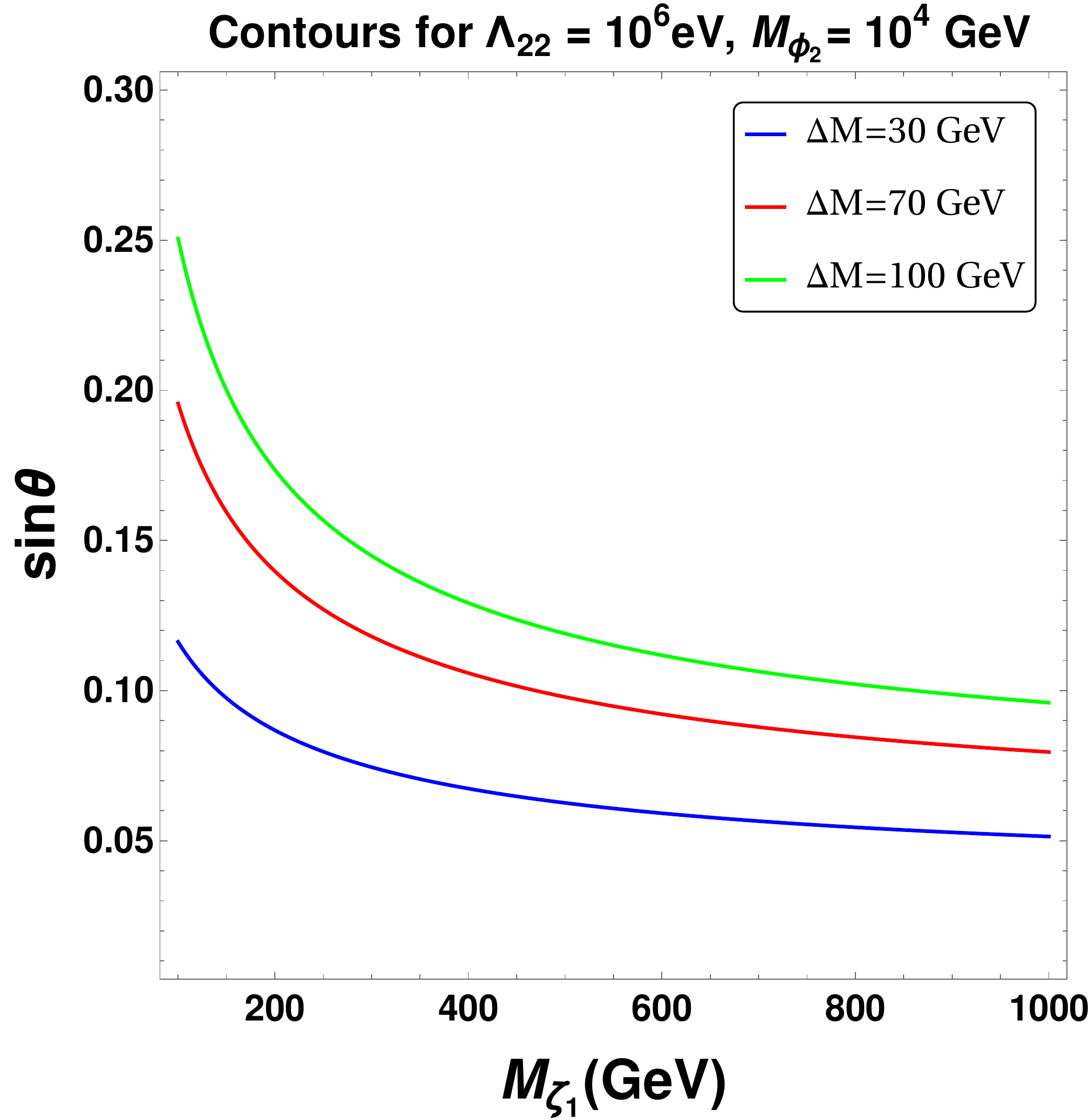}~
\includegraphics[scale=0.35]{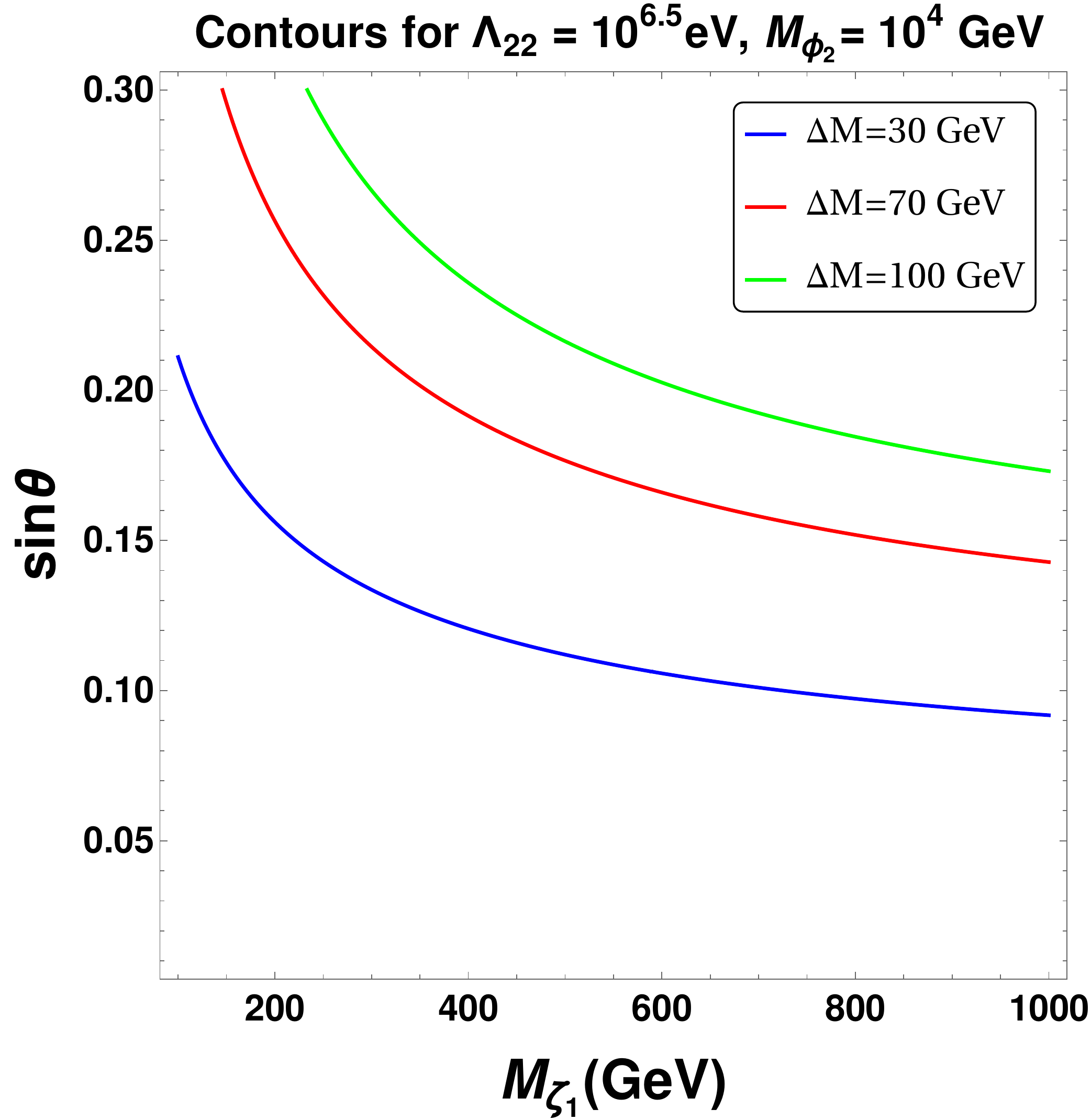}
\caption{(Upper plots) demonstrate the contours for $\Lambda_{11}$ for different values of $\Delta M$ in $\sin \theta - M_{\zeta_1}$ plane.
Similarly, (lower plots) demonstrate Contours for $\Lambda_{22}$. 
}
\label{lambda}
\end{figure}
%========================================

In Fig.~\ref{lambda} (upper plots), we present the contours for $\Lambda_{11} = 10^5$~eV (left panel), $\Lambda_{11} = 10^{5.5}$~eV (right panel)  considering several values of $\Delta M$ in the $\sin \theta - M_{\zeta_1}$ plane. For this purpose, we fix $m_{\chi_{L,R}}= 1$~GeV and $M_{\phi_1}$ at $1.2 \times 10^{3}$~GeV. It is evident from this figure that, for a necessity of higher values of $\Lambda_{11}$ one has to go for larger  $\sin\theta$ values.  
In Fig.~\ref{lambda} (lower plots), we present the contours for $\Lambda_{22} = 10^{6}$~eV (left panel), $\Lambda_{22} = 10^{6.5}$~eV (right panel) considering the set of earlier values of $\Delta M$ in the $\sin \theta - M_{\zeta_1}$ plane. Here also we take $m_{\chi_{L,R}} = 1$~GeV and fix  $M_{\phi_2}$ at $10^{4}$~GeV. One can draw a similar conclusion on the contours of $\Lambda_{22}$ as we get for $\Lambda_{11}$.

It is to note that, in order to make the three SM neutrinos massive one needs to take the presence of three scalars, although it is sufficient to have two scalars only for a scenario where one of the active neutrinos remains massless. In the presence of a third copy of the scalar, we would have evaluated the corresponding $\Lambda$ in a similar manner.

Once we construct the light neutrino mass matrix with the help of different $\Lambda_{ij}$s we can study the properties associated with neutrino mass. The obtained low energy neutrino mass matrix $m_{\nu_{ij}}$ thus constructed is diagonalized by the unitary matrix $U_{\nu} (U)$.
\begin{align}
       m_{\nu}^{\rm diag}=U^T m_\nu U,
       \end{align}
 We consider the charged lepton matrix to be diagonal in this model. In that case, we can identify $U$ as the standard $U_{\rm PMNS}$ matrix~\cite{Maki:1962mu} for lepton mixing. 
 
 %========================================
\begin{table}[t]
\begin{center}
\begin{tabular}{| c | c | c | c | c | c | c | c | c|}
  \hline
  SL no. & $M_{\zeta_1} $ (GeV) & $\Delta M$ (GeV) & $\sin\theta$ &  $\Omega h^2$ & $ {\rm Log}_{10}\left[\frac{\sigma ^{\rm SI}}{\text{cm}^2}\right]$ &$\Lambda_{11}$ (eV)  & $\Lambda_{22}$ (eV) & $\Lambda_{33}$ (eV) \\
  \hline
  I & 200 &  47  & 0.256  &0.12 & -46.71 & $1.95\times 10^{6}$ & $5.04\times 10^{6}$  & $8.44\times 10^{6}$ \\
  \hline
  II & 800 &  123  & 0.066 & 0.12 & -48.26 & $2.79\times 10^{5}$ & $3.38\times 10^{5}$ & $7.18\times 10^{5}$ \\
  \hline
\end{tabular}
\caption{Two sets of relic and direct search satisfied points and corresponding values of $\Lambda$ considering $m_{\chi_{L,R}}\sim $~1 GeV, scalar field masses,  $M_{\phi_i}\sim\{1.2 \times 10^{3},~10^{4},10^{5}\}$ (GeV) and the lightest active neutrino mass $m_\nu^{\rm lightest}\sim 0.01$ eV. The points are also tested to satisfy Br$(\mu \rightarrow e \gamma)$ bound.}
\label{tab:tab2}
\end{center} 
\end{table}
%======================================== 

 To start with Eq.~(\ref{mnu}), one can get the light neutrino mass in terms of the Yukawa couplings $h_{ij}$ and the mass scale $\Lambda_{kk}$. The $h_{ij}$ which is present in Eq.~(\ref{mnu}) can be connected to the oscillation parameters with the help of Casas-Ibarra parameterization~\cite{Casas:2001sr}, which allows us to use a random complex orthogonal rotation matrix $\mathcal{R}$. Using this parameterization, we can express the Yukawa coupling by the following equation~\cite{Casas:2001sr}.
 \begin{equation}
  h^T = D_{\!\!\sqrt{\Lambda^{-1}}} \, \mathcal{R} \, D_{\!\!\!\!\sqrt{m_\nu^{\rm diag}}} \, U^{\dagger},
 \end{equation}
\noindent where, $D_{\!\!\!\!\sqrt{m_\nu^{\rm diag}}} = {\rm Diag}(\sqrt{m_{\nu 1}},~\sqrt{m_{\nu_ 2}},~\sqrt{m_{\nu 3}}), ~~ D_{\!\!\sqrt{\Lambda^{-1}}} = {\rm Diag}(\sqrt{\Lambda_{11}^{-1}},~\sqrt{\Lambda_{22}^{-1}},~\sqrt{\Lambda_{33}^{-1}})$.
The $\mathcal{R}$ can be parameterised through three arbitrary mixing angles which we choose to be $(\frac{\pi}{4},\frac{\pi}{3},~{\rm and}~\frac{\pi}{6})$. Now 
to have a numerical estimate of the Yukawa couplings $h_{ij}$, as stated earlier we consider $m_{\chi_{L,R}}$ at 1 GeV  and scalar field masses at $\{1.2 \times 10^{3},~10^{4},10^{5}\}$ GeV and make use of two sets of relic density and direct search satisfied points as tabulated in Table~\ref{tab:tab2}. At the same time, we use best fit central values of the oscillation parameters to construct the $U_{\rm PMNS}$ matrix and choose the normal hierarchy mass pattern~\cite{Tanabashi:2018oca} with the lightest active neutrino mass eigenvalue as 0.01 eV. In Table~\ref{tab:tab3} we represent the Yukawa coupling matrices ($h$) using the above sets of benchmark points.
So far, the analysis of neutrino part has been carried out by keeping $m_\chi$ fixed at 1 GeV. One can go for an even higher choice of
$m_{\chi_{L,R}}$ values (competent with the pseudo-Dirac limit), however, in such a scenario the order of the elements of the $h$ matrix will be reduced further as evident from Eq.~(\ref{mnu}). One can choose arbitrary masses for the scalars for generating the active neutrino mass radiatively at one loop order as described before. However corresponding Yukawas $h_{ij}$ would be suitably modified such that higher values in  $M_{\phi_i}$s would suppress them further than our benchmark scenario, represented in Table \ref{tab:tab3}.
%In a similar way, $M_{\phi_i}$s can take larger values as well compared to our choices in Table \ref{tab:tab3} which would again lead to reduced Yukawas $h_{ij}$.

%========================================
 \begin{table}[t]
\begin{center}
\begin{tabular}{| c | c |}
  \hline
  SL no. & $h_{ij}$ \\
  \hline
  I &  $10^{-5}\times\left(
\begin{array}{ccc}
 -4.26+2.29 i & 2.38\, -1.01 i & -2.03-0.75 i \\
 2.67\, -2.09 i & 3.10\, -4.42 i & 3.51\, -2.60 i \\
 7.44\, -7.15 i & 3.29\, -2.30 i & -0.076-1.03 i \\
\end{array}
\right)$\\
  \hline
  II & $10^{-4}\times\left(
\begin{array}{ccc}
 -1.13+0.60 i & 0.92\, -0.39 i & -0.70-0.26 i \\
 0.71\, -0.55 i & 1.20\, -1.70 i & 1.20\, -0.90 i \\
 1.97\, -1.90 i & 1.27\, -0.89i & -0.026-0.35 i \\
\end{array}
\right)$ \\
  \hline
\end{tabular}
\caption{Numerical estimate of the two Yukawa coupling matrices which are built for the sets of benchmark points tabulated in Table~\ref{tab:tab2}.}
\label{tab:tab3}
\end{center}
\end{table}
%========================================

It is expected that constraint on the model parameter, specifically $h_{ij}$ may arise from the lepton flavour–violating (LFV) decays of $\phi$ fields. The most stringent limit
comes from the $\mu\rightarrow e\gamma$ decay process~\cite{Dinh:2012bp,Tommasini:1995ii,Ilakovac:1994kj}. However, the Yukawa couplings being very small $\sim \mathcal{O}(10^{-5})$ as tabulated in Table \ref{tab:tab3} easily overcome the present  experimental bound~\cite{Baldini:2018nnn}.
The pseudo-Dirac nature of dark matter is testable at colliders through displaced vertices~\cite{Davoli:2017swj}.  A detailed study is required whether a relaxed $\sin \theta$ has some role to play in this regard. Constraints on the model parameter are under consideration~\cite{WorkCont}.

%%%%%%%%%%%%%%%%%%%%%%%%%%%%%%%%%%%%%%%%%%%%%%%%%%%%%%%%%%%%%%%%%%%%%
\section{Conclusion}\label{conclusion}
%%%%%%%%%%%%%%%%%%%%%%%%%%%%%%%%%%%%%%%%%%%%%%%%%%%%%%%%%%%%%%%%%%%%%
In this work, we study a simple extension of the standard model, including a singlet doublet dark sector in the presence of a small Majorana mass term. As a consequence generated eigenstates deviate from Dirac nature, owing to a small mass splitting between pair of two pseudo-Dirac states. Lightest of these pseudo-Dirac fermionic states,  considered as dark matter, can evade the strong spin-independent direct detection constrain by suppressing the scattering of dark matter with nucleon through the Z-boson mediation. We explicitly demonstrate this significant weakening of the direct detection constraint on the singlet doublet mixing parameter while ensuring that such dark matter is still capable of satisfying the thermal relic fully.
 
The same Majorana mass term provides an elegant scope to generate neutrino mass radiatively at one loop, which requires an extension of the dark sector model with copies of real scalar singlet fields.  Introduction of these additional scalars is also motivated by stabilizing the electroweak vacuum even in the presence of a large mixing angle. They also provide a source of lepton number violation, generating light Majorana neutrinos satisfying oscillation data fully.  Hence this present scenario offers the potential existence of a pseudo-Dirac type dark matter in the same frame with light Majorana neutrinos. 
We obtain two different bounds on the left and right component of the newly introduced Majorana mass parameter, {\sl i.e.} ($m_{\chi_L} + m_{\chi_R}) \gtrsim \mathcal{O}(1) $~GeV and ($m_{\chi_L} - m_{\chi_R}) \lesssim \mathcal{O}(1) $~MeV, accounting for the correct order of active neutrino masses and oscillation data. We further demonstrate the dependence of these model parameters and reference benchmark points satisfying best fit central values of the oscillation parameters and consistent with the pseudo-Dirac dark matter constraints.

%%%%%%%%%%%%%%%%%%%%%%%%%%%%%%%%%%%%%%%%%%%%%%%%%%%%%%%%%%%%%%%%%%%%%
\section*{Acknowledgements}\label{ack}
%%%%%%%%%%%%%%%%%%%%%%%%%%%%%%%%%%%%%%%%%%%%%%%%%%%%%%%%%%%%%%%%%%%%%
This work is supported by Physical Research Laboratory (PRL), Department of Space, Government of India. Computations were performed using the HPC resources (Vikram-100 HPC) and TDP project at PRL. Authors gratefully acknowledge WHEPP'19 where parts of this work were initiated. Authors also thank KM Patel and S Seth for useful discussion.

\newpage
%%%%%%%%%%%%%%%%%%%%%%%%%%
%\bibliographystyle{apsrev4-1}
\bibliographystyle{JHEP}
\bibliography{refs}
%%%%%%%%%%%%%%%%%%%%%%%%%%
\end{document}